\documentclass[runningheads]{llncs}
\usepackage{graphicx}
\usepackage{booktabs}
\usepackage{hyperref}
\usepackage{multirow}
\begin{document}
\title{HypeRS: Building a Hypergraph-driven ensemble Recommender System} 
%
\author{Alireza Gharahighehi\inst{1,2} \and Celine Vens \inst{1,2} \and
Konstantinos Pliakos\inst{3}}

%
\authorrunning{A. Gharahighehi et al.}
%

\institute{Itec, imec research group at KU Leuven, Kortrijk, Belgium \and
KU Leuven, Campus KULAK, Department of Public Health and Primary Care, Kortrijk, Belgium \and KU Leuven, Department of Management, Strategy and Innovation, Leuven, Belgium
\email{\{alireza.gharahighehi,celine.vens,konstantinos.pliakos\}@kuleuven.be}
}
\maketitle              
\begin{abstract}
Recommender systems are designed to predict user preferences over collections of items. These systems process users' previous interactions to decide which items should be ranked higher to satisfy their desires. An ensemble recommender system can achieve great recommendation performance by effectively combining the decisions generated by individual models. In this paper, we propose a novel ensemble recommender system that combines predictions made by different models into a unified hypergraph ranking framework. This is the first time that hypergraph ranking has been employed to model an ensemble of recommender systems. Hypergraphs are generalizations of graphs where multiple vertices can be connected via hyperedges, efficiently modeling high-order relations. We differentiate real and predicted connections between users and items by assigning different hyperedge weights to individual recommender systems. We perform experiments using four datasets from the fields of movie, music and news media recommendation. The obtained results show that the ensemble hypergraph ranking method generates more accurate recommendations compared to the individual models and a weighted hybrid approach. The assignment of different hyperedge weights to the ensemble hypergraph further improves the performance compared to a setting with identical hyperedge weights.
\end{abstract}

\section{Introduction}
Nowadays, people use digital services more and more to fulfill their needs. The owners of these services monitor users' behavior and utilize users' interactions with provided items, such as movies, songs, commercial products, to predict users' preferences. This enables the personalization of digital services and the rise of effective recommender systems (RSs) which learn from users' preferences and provide them with accurate recommendations. Generally, there are two main categories in RSs: content-based filtering and collaborative filtering approaches. Content-based RSs use the features that describe the items for computing similarities between the items and the user interaction profile. Next, they  recommend items that are more similar to this user profile. Upon a recommendation query for a target user, these RSs do not consider the interactions of the other users in generating the recommendation list. In contrast to that, collaborative filtering approaches infer the users' preferences by processing the collaborative information between users or items. In many applications, collaborative filtering RSs generate more accurate~\cite{adomavicius2005toward} and less obvious~\cite{Lops2011} recommendations compared to content-based approaches.

Each type of RS processes the information based on different assumptions to decide which items should be ranked higher among many available ones. For instance, memory-based collaborative filtering approaches (user-based and item-based) assume that users (items) with similar interactions have similar interests. Therefore, these approaches form neighborhoods to generate recommendations. Model-based collaborative filtering approaches assume that users and items can be represented in a common feature space and they use different learning methods to learn these latent features. While these approaches might vary in prediction power, they convey relevant information from different perspectives, following practically different learning strategies for the same recommendation task. Ensemble methods include multiple learning methods and integrate their predictive power into a single system, achieving superior predictive performance to individual models. Examples of ensembles in machine learning are bagging and boosting. In recommendation tasks a hybrid RS can be applied to exploit several data sources or the prediction power of different RSs to generate more relevant recommendations. An ensemble RS is a hybrid model that employs the ranking lists of multiple RSs to decide which items should be recommended to each user~\cite{Aggarwal2016}.

In this paper we propose an ensemble hypergraph learning framework for recommendation. This way we integrate the predictive power of several models into a unified RS powered by hypergraph ranking. Unlike regular graphs, where edges connect pairs of nodes, in hypergraphs multiple nodes can be connected via hyperedges. These higher order relations in hyperedges empower hypergraphs to cast more reliable information in the model~\cite{vinayak2016crowdsourced}. Furthermore, hypergraph learning can inherently model the complex relations between different types of entities in a unified framework. It is therefore a deliberate choice for the construction of an ensemble of individual RSs driven by different types of information. Moreover, as was shown in \cite{fairHyper}, hypergraph ranking-based methods can mitigate popularity bias, enhance fairness and coverage as well as act as innate multi-stakeholder RSs. The main contribution of this paper is to construct a hypergraph as an ensemble framework for recommendation tasks. Despite its capability to stack multiple connections in a unified model, to the best of our knowledge hypergraphs have not been employed to form ensembles of RSs. 

This article is an extension of our previous conference paper~\cite{gharahighehi2021ensemble} in the following main directions:

\begin{itemize}
    \item We have extended the hypergraph-based ensemble RS by further incorporating two effective individual RSs, namely Weighted Approximate-Rank Pairwise (WARP)~\cite{weston2011wsabie} and Multi-variational autoencoder (MVAE)~\cite{liang2018variational}. This way, we managed to substantially enhance the performance of our RS. We also demonstrated the potential of our model, as it has the capacity to integrate more effective individual RSs that might appear in the future, boosting its recommendation performance even further.

    \item In the previous paper we considered equal weights for all the hyperedges in the ensemble hypergraph. Here, we have  differentiated the actual links between users and items from the ones predicted by the individual RSs. We also assigned different weights to hyperedges that relate to different individual RSs based on the performance of these RSs.
    
    \item We have extended the experimental study by including more evaluation metrics as well as by comparing our method to two additional baseline models, namely Weighted Approximate-Rank Pairwise (WARP)~\cite{weston2011wsabie} and Multi-variational autoencoder (MVAE)~\cite{liang2018variational}. Especially the MVAE is a state-of-the-art method that emerged among the winning models in a recent comparison study of RSs \cite{FerrariRecSys:2019}.

\end{itemize}

The structure of this paper is as follows: Studies about applications of hypergraph learning in RSs are presented in Section~\ref{sec2}. Next, in Section~\ref{sec3}, we show how a unified hypergraph can be formed as an RS (Section~\ref{sec3_1}) and how it can formulate an ensemble of RSs (Section~\ref{method_ensemble}). In Section~\ref{sec4}, four recommendation datasets are described and the experimental setup in designing and testing the proposed model is described. Next, the obtained results of comparing the proposed ensemble model against other methods on these four datasets are presented and discussed in Section~\ref{sec5}. Finally, we draw conclusions and outline some directions for future research in Section~\ref{sec6}.  

\section{Related work}
\label{sec2}
Hypergraph learning has been applied to generate recommendation lists in several applications. For instance in the music domain, Bu et al.~\cite{bu2010music} used hypergraph learning to recommend music tracks where the relations between users, tracks, albums and artists were modeled using a unified hypergraph. Hypergraph ranking has been also used in news recommendation tasks~\cite{li2013news,fairHyper}. News usually contains very rich features such as text, tags and named entities. Therefore, hypergraph learning can effectively model the relations between these entities. Moreover, Pliakos et al.~\cite{pliakos2014simultaneous} used hypergraph ranking for a tag recommendation task. They built a hypergraph ranking model to capture the complex relations between different entities in the system, such as users, images, tags, and geo-tags. Hypergraph-based RSs have been also used in e-commerce applications~\cite{Mao:2019,Wang:2020}. For instance in~\cite{Mao:2019}, a multipartite hypergraph is used to model the relations between users, restaurants and attributes in a multi-objective setting. In such applications, item attributes and sequences of user-item interactions are effectively modeled in hypergraphs.


Hypergraph learning has been employed to address various issues in RSs. A hypergraph can model the relations between different types of stakeholders and objects and therefore, it can be intrinsically used as a multi-stakeholder RS~\cite{fairHyper}. Additionally, it can be used to burst the filter bubble around the user by querying a more diverse recommendation list based on the user history~\cite{li2013news,fairHyper}. Moreover, hypergraph learning has been used to address fairness~\cite{fairHyper}, the cold-start problem~\cite{zheng2018novel} as well as context-awareness~\cite{yu2018modeling} in recommendation tasks. 


An ensemble RS is a type of hybrid RSs that integrates the recommendations of multiple individual RSs. Aggarwal~\cite{Aggarwal2016} categorized hybrid RSs to monolithic, ensembles, and mixed RSs. Burke et al.~\cite{burke2002hybrid} provided another categorization where hybrid models are categorized into weighted, switching, cascade, feature augmentation, feature combination, meta-level and mixed RSs. A weighted hybrid RS uses the weighted average of the scores from individual RSs to generate the recommendation list. For instance, Do et al.~\cite{9061465} applied a weighted hybrid RS based on collaborative and content-based filtering approaches on \textit{Movielens} dataset and showed that it is more effective compared to the individual collaborative and content-based RSs. Here, we employ a unified hypergraph as an ensemble RS. Although hypergraph learning is very promising and effective in addressing many problems in RSs, to the best of our knowledge, it has never been studied as an ensemble RS.

\section{Methodology}
\label{sec3}
\subsection{Hypergraphs as recommender systems}
\label{sec3_1}

Hereafter, uppercase bold letters are used for matrices, lowercase bold letters represent vectors, uppercase non-bold letters are used for sets and lowercase non-bold letters represent constants. The element in $i^{th}$ row and $j^{th}$ column of matrix \textbf{X} is denoted as \textbf{X}(i,j). 

A hypergraph consists of a set of nodes (vertices) $N: \{n_1,n_2\cdots,n_{|N|}\}$ and a set of hyperedges $E: \{e_1,e_2\cdots,e_{|E|}\}$ that connect the nodes. Each hyperedge can connect multiple nodes in the hypergraph. Based on the application, different types of hyperedges can be defined that capture different forms/sources of information. We define these hyperedge types in Section~\ref{method_ensemble}. In a typical collaborative filtering setting there are two types of entities in a hypergraph: users $U: \{u_1,u_2\cdots,u_{|U|}\}$ and items $I: \{i_1,i_2\cdots,i_{|I|}\}$. Therefore, the set of nodes $N$ in a hypergraph is formed based on users and items ($N:  \{U\cup I\}$).

Let $\mathbf{H}$ of size $|N|\times|E|$ be the incidence matrix of the hypergraph, where $H(n,e)=1$, if node $n$ is in hyperedge $e$ and \textit{zero} otherwise. Based on $\mathbf{H}$, the symmetric matrix $\mathbf{A}$ can be formed using Eq.\ref{eq1}:

\begin{equation}
\mathbf{A}=\mathbf{D_{n}}^{-1/2}\mathbf{H}\mathbf{W}\mathbf{D_e}^{-1}\mathbf{H}^{T}\mathbf{D_{n}}^{-1/2}
\label{eq1}
\end{equation}

\noindent where $\mathbf{D}_{n}$ and $\mathbf{D}_{e}$ are the diagonal matrices that contain the node and hyperedge degrees and $\mathbf{W}$ is the diagonal hyperedge weight matrix. Each element $\mathbf{A}(i,j)$ reflects the relatedness between nodes $i$ and $j$. Higher values indicate stronger relations between the corresponding nodes. Then, the recommendation problem is formulated as finding a ranking (score) vector $\mathbf{f}\in {\rm I\!R}^{|N|}$ that minimizes the following loss function~\cite{bu2010music}:

\begin{equation}
Q(\mathbf{f})=\frac{1}{2}\mathbf{f}^{T} \mathbf{L} \mathbf{f}+\vartheta||\mathbf{f} - \mathbf{y}||_2^2
\label{eq2}
\end{equation}

\noindent where $\mathbf{L}$ is the hypergraph Laplacian matrix (i.e. $\mathbf{L}=\mathbf{I}-\mathbf{A}$), $\vartheta$ is a regularizing parameter and $\mathbf{y} \in {\rm I\!R}^{|N|}$ is the query vector. Every item of the ranking vector $\mathbf{f}$ or query vector $\mathbf{y}$ corresponds to a node. Typically, to generate the recommendation list for user \(u\) in a regular recommendation task, one can query the hypergraph for user $u$ by setting the corresponding value in the query vector to \textit{one} ($\mathbf{y}(u)=1$) and all the other values that correspond to other nodes to \textit{zero}. By solving the optimization problem in Eq.\ref{eq2}, the optimal score (ranking) vector can be calculated using Eq.\ref{eq3}:

\begin{equation}
{\mathbf{f}}^{\ast} = \frac{\vartheta}{1+\vartheta}\bigl(\mathbf{I} - \frac{1}{1+\vartheta} \mathbf{A} \bigr)^{-1}\mathbf{y}.
\label{eq3}
\end{equation}
Finally, the top k items that have the highest scores in $\mathbf{{f}^{\ast}}$ are recommended to the user $u$. 

\subsection{A hypergraph-based ensemble recommender system, HypeRS}
\label{method_ensemble}

An ensemble RS\footnote{The source code is available at \url{https://github.com/alirezagharahi/ensemble\_hypergraph}.} utilizes the decisions of multiple individual RSs to decide which items should be ranked higher in the final recommendation lists. Let $M: {\{m_1,m_2\cdots,m_{|M|}\}}$ be the set of individual methods that we want to incorporate in our ensemble RS. Each of these individual methods $m_{i}$ can generate its own top \(k\) rankings $\mathbf{R}_{i}\in {\rm I\!R} ^{|U|\times k}$ where each row in $\mathbf{R}_{i}$ is the top \(k\) ranked items for the corresponding user. Then, based on the recommendation lists of each RS, hyperedges are formed to connect users to their top \(k\) recommendations. 

As is mentioned previously, the hypergraph consists of multiple types of hyperedges. We consider three types of hyperedges, which are defined in Table~\ref{tab:edges}. The $E_{UI}$ hyperedges connect the users with the items that they have interacted with. To make the relations between users with similar tastes more explicit, the $E_{UU}$ hyperedges connect users to their \textit{k} nearest neighbors. To find these neighbors we use the user-item interaction matrix $\mathbf{Z}$, where $\mathbf{Z}(i,j)\in \{0,1\}$. The \textit{k} nearest neighbors of user \textit{u} are users that have the highest cosine similarity with \(u^{th}\) row of matrix $\mathbf{Z}$. The $E_{M}$ hyperedges are considered to integrate the recommendations of multiple RSs in the hypergraph. These RSs can be from different families such as collaborative filtering or content-based approaches. The fact that recommendations from any type of RS can be directly modeled as hyperedges in our system is a vital advantage of the proposed method. 

We constructed the $E_M$ hyperedge set using four well-established and powerful matrix completion-based recommendation methods, namely BPR~\cite{rendle2012bpr}, WARP \cite{weston2011wsabie}, WRMF~\cite{hu2008collaborative,pan2008one} and MVAE~\cite{liang2018variational}. \textit{BPR} is a learning-to-rank matrix completion approach which uses user-specific relative pair-wise preferences between observed and unobserved items to learn items' and users' low rank matrices. Similar to \textit{BPR}, \textit{WARP} is also a pair-wise learning-to-rank approach but with a different objective function. \textit{BPR} is optimized for approximation of \textit{AUC} whereas \textit{WARP} is optimized for \textit{precision}. \textit{WRMF} is a matrix factorization approach for implicit feedback datasets that uses the alternating-least-squares optimization process to learn items' and users' parameters. MVAE is a CF model for implicit feedback that assumes that user logs are from a  multinomial distribution and uses variational autoencoders to learn users' and items' parameters.

\begin{table*}[h]
\centering
  \caption{Hyperedge definitions}
  \label{tab:edges}
  \begin{tabular}{p{0.15\textwidth}p{0.45\textwidth}c}
    \toprule
    \textbf{hyperedge}&\textbf{definition}&\textbf{\# of hyperedges}\\
    \midrule
    \ $E_{UI}$ & Each user is connected to the items that the user has interacted with & $|U|$\\
    \ $E_{UU}$ & Each user is connected to the k most similar users & $|U|$\\
    \ $E_{M}$ & Each user is connected to top k recommended items by a RS & $|M|\times|U|$\\
   \bottomrule
\end{tabular}
\end{table*}

The hypergraph and its incidence matrix $\mathbf{H}$ are constructed using the hyperedge sets of Table \ref{tab:edges}. Following that, the affinity matrix $\mathbf{A}$ is computed and the recommendation task is addressed as was described in Section \ref{sec3_1}.

The weight of a hyperedge reflects the relative importance of the hyperedge compared to the other hyperedges in the hypergraph. We propose to assign lower weights for $E_{M}$ hyperedges (0.5) compared to $E_{UI}$ ones (1.0), as the $E_{M}$ are based on predicted links, whereas the $E_{UI}$ are based on real links between users and items. One should also consider that the individual RSs in the ensemble have different predictive power. Therefore we assign a decay weight based on their performance in the validation set. Hyperedges that are associated with the top ranked RS receive no decay weight while hyperedges that are related to lower ranked RSs receive lower weights (according to their ranks). In this paper we consider a linear decay (10\%) per rank.      


\section{Experimental Setup}
\label{sec4}
To evaluate the performance of the proposed approach we use four datasets from news, music and movie application domains. These datasets are described in Table~\ref{tab:datasets}. AOTM is a publicly available dataset collected from the Art-of-the-Mix platform that is based on user playlists~\cite{mcfee2012hypergraph}. Movielens\footnote{\url{http://www.grouplens.org}} is a publicly available movie rating dataset ~\cite{Cantador:RecSys2011}. As we only encode interactions in the hypergraph for this dataset we transform ratings to binary feedback. Globo\footnote{\url{http://www.globo.com}} and Roularta\footnote{\url{http://www.roularta.be}} are news datasets that contain readers' interactions with news articles.  

\begin{table*}[h]
\centering
  \caption{Datasets descriptions}
    \label{tab:datasets} 
   
  \begin{tabular}{lcccc}
    \toprule
    & \textbf{AOTM}&\textbf{Movielens}&\textbf{Globo}&\textbf{Roularta}\\
    \midrule
    \textbf{item type}&music track&movie&news article&news article\\
    \textbf{\# users} &1,605&1,573&3,903&5.082\\
    \textbf{\# items}&2,199&2,053&1,246&2,739\\
    \textbf{sparsity} &3.8\%&19.9\%&5.7\%&8.5\%\\
  \bottomrule
\end{tabular}
\end{table*}

In our experiments we consider the following eight approaches\footnote{For BPR and WRMF we used \hyperlink{https://implicit.readthedocs.io/en/latest/index.html}{implicit library}.}:

\begin{itemize}
\item \textbf{BPR}: Bayesian Personalized Ranking (BPR)~\cite{rendle2012bpr} is a pair-wise learning-to-rank matrix completion approach as presented in the previous section.   

\item \textbf{WARP}: Weighted Approximate-Rank Pairwise (WARP)~\cite{weston2011wsabie} is a another pair-wise learning-to-rank method as explained in the previous section.

\item \textbf{WRMF}: Weighted Regularized Matrix Factorization (WRMF)~\cite{hu2008collaborative,pan2008one} is a MF approach using the alternating-least-squares optimization process to learn items and users' parameters as presented in the previous section.

\item \textbf{MVAE}: Multi-variational autoencoder (MVAE)~\cite{liang2018variational} is a CF approach based on variational autoencoders as explained in the previous section. 

\item \textbf{Hybrid}: A weighted hybrid model that uses scores of \textit{BPR}, \textit{WARP}, \textit{MVAE} and \textit{WRMF}, and then considers the weighted average of these scores to generate the final ranking lists.

\item \textbf{H}: A hypergraph-based RS explained in Section~\ref{sec3_1} that only contains the hyperedge types of $E_{UI}$ and $E_{UU}$ from Table~\ref{tab:edges}.
\item $\mathbf{HypeRS}$: The proposed hypergraph-based ensemble RS explained in Section~\ref{method_ensemble}.
\item $\mathbf{HypeRS_{W}}$: The proposed hypergraph-based ensemble RS with proposed hyperedge weights.
\end{itemize}

To validate the performance of the proposed method against the compared methods we randomly hide \textit{ten} interactions of each user from training and then measure the ability of the methods in predicting these hidden interactions~\footnote{Users with few interactions are omitted from experiments.}. We use \textit{precision@10}, \textit{recall@10} and \textit{F1-score@10} to measure the accuracy of predictions. \textit{Precision} and \textit{Recall} are standard information retrieval accuracy measures that reflect the proportion of the recommendation list which is relevant and the proportion of relevant items that is recommended respectively. To have a balanced measure, one can calculate \textit{F1-score} that combines the precision and recall into a single measure by taking their harmonic mean. These accuracy measures can be calculated as follows:

\begin{equation}
Precision = \frac{1}{|U|}\sum_{u \in U} \frac{|T_{u} \cap R_{u}|}{|R_{u}|},
\label{eq:B_precision}
\end{equation}
\begin{equation}
Recall = \frac{1}{|U|}\sum_{u \in U} \frac{|T_{u} \cap R_{u}|}{|T_{u}|}.
\label{eq:B_recall}
\end{equation}

\begin{equation}
F1-score = 2 \times \frac{precision \times recall}{precision + recall}.
\label{eq:B_f1}
\end{equation}

\noindent where $U$ is the set of users, $T_{u}$ is the test items for user $u$, and $R_{u}$ is the recommendation list for user $u$. As the number of relevant items and the length of the recommendation lists are the same (10 items), the reported results from \textit{precision}, \textit{recall} and \textit{F1-score} are eventually all the same. 

The compared methods have some hyperparameters to be tuned. \textit{BPR} and \textit{WARP} have number of latent features, number of iterations, regularizing parameter and learning rate, \textit{WRMF} has number of latent features, number of iterations, regularizing parameter, \textit{MVAE} has batch size, number of iterations and number of anneal steps, \textit{Hybrid} model has hybridization weights and \textit{H} as well as $HypeRS$ have a regularizer as a hyperparameter. To tune these hyperparameters we form a validation set for each dataset by randomly drawing  \textit{five} interactions of each user from the training set as the validation set. The final tuned hyperparameter values are based on \textit{precision@10} and are reported in Table~\ref{tab:parameters}. 

\begin{table}[h]
\centering
  \caption{Hyperparameters}
    \label{tab:parameters} 
   
  \begin{tabular}{llccccc}
    \toprule
    && range&\textbf{AOTM}&\textbf{Movielens}&\textbf{Globo}&\textbf{Roularta}\\
    \midrule
    \multirow{4}{*}{BPR}
    &\# iterations& [1000,2000]& 1645 & 1984 & 1598 & 1984\\
    &\# latent features& [100,250] & 179 & 129 & 168 & 129\\
    & regularizing parameter& [0.01,0.05]& 0.0194 & 0.0412 & 0.0374&0.0412\\
    & learning rate &[0.001,0.07]& 0.0284 & 0.0092 & 0.0174&0.0092\\
    \midrule
    \multirow{4}{*}{WARP}
    &\# iterations& [200, 850]& 809 & 810 & 810 & 650\\
    &\# latent features& [15,40] & 21 & 21 & 21 & 26\\
    & regularizing parameter& [10e-3,10e-6]& 4.7e-06 & 3.4e-6 & 8.1e-6&7.4e-6\\
    & learning rate &[0.001,0.1]& 0.0472 & 0.0165 & 0.0191&0.0191\\
    \midrule
    \multirow{3}{*}{WRMF}
    &\# iterations & [1000,2000]&1276 & 1393 & 29 & 1288\\
    &\# latent features & [100,250]&201 & 107 & 493 & 109\\
    & regularizing parameter & [0.01,0.05]&0.0374 & 0.0225 & 0.0432&0.0315\\
    \midrule
    \multirow{3}{*}{MVAE}
    &\# iterations & [10,250]&29 & 18 & 29 & 28\\
    & batch size & [25,500]&469 & 34 & 152 & 127\\
    &\# anneal steps & [100000,300000]&127692 & 100212 & 244065&223544\\
    \midrule
    \multirow{4}{*}{Hybrid}
    &BPR\_weight&[0.01,0.99]&0.181&0.152&0.443&0.285\\
    &WARP\_weight&[0.01,0.99]&0.422&0.193&0.290&0.184\\
    &WRMF\_weight&[0.01,0.99]&0.0588&0.490&0.458&0.168\\
    &MVAE\_weight&[0.01,0.99]&0.329&0.255&0.469&0.427\\
    \midrule
    H &regularizing parameter&[0.01,0.99]&0.2414&0.2414&0.0656&0.0616\\
    \midrule
    $HypeRS$&regularizing parameter&[0.01,0.99]&0.4554&0.4554&0.8301&0.6325\\
  \bottomrule
\end{tabular}
\end{table}

\section{Results and Discussion}
\label{sec5}
The results of the proposed hypergraph-based ensemble RS and the selected approaches on the four datasets are reported in Table~\ref{tab:results}. The reported values are in terms of average  \textit{precision@10} of the recommendation lists generated by the compared approaches. As is shown in Table~\ref{tab:results}, the proposed hypergraph-based ensemble RS (\(HypeRS\)) has superior predictive performance compared to all the competitor approaches including the weighted hybrid model in all datasets. The competitor methods have different performance rankings in the four datasets. Each of these methods processes the information based on different assumptions and learning approaches. The effectiveness of these assumptions and learning approaches differs across different applications. An ensemble RS exploits the combined predictive power of the individual methods. It considers all assumptions and decisions of various independent RSs and achieves overall superior performance regardless of the application domain of the recommendation task. The weighted ensemble RS (\(HypeRS_{W}\)) performs better compared to the ensemble RS with identical hyperedge weights (\(HypeRS\)) in all datasets. 

\begin{table*}[h]
\centering
  \caption{Results w.r.t. \textit{precision@10/recall@10/F1-score@10}}
    \label{tab:results} 
  \begin{tabular}{lcccc}
    \toprule
    & \textbf{AOTM}&\textbf{Movielens}&\textbf{Globo}&\textbf{Roularta}\\
    \midrule
BPR      & 0.0373 & 0.1716    & 0.0937 & 0.0704 \\
WRMF      & 0.0402 & 0.1718    & 0.0921 & 0.0918\\
MVAE &0.0426&0.1690&0.1487&0.0837\\
WARP&0.0360&0.1870&0.1375&0.0907\\
H        & 0.0338 & 0.1503    & 0.1125 & 0.0657\\
Hybrid & 0.0412 & 0.1861    & 0.1012 & 0.0944\\

$HypeRS$ & 0.0479 & 0.1957& 0.1499 & 0.0950\\
$HypeRS_{W}$ &\textbf{0.0483}&\textbf{0.1995}&\textbf{0.1504}&\textbf{0.0955}\\

  \bottomrule
\end{tabular}
\end{table*}

In this study we keep the experiments simple by only using the collaborative information, i.e. user-item interactions, to make them applicable on available datasets and various application fields (i.e. movies, music, news). Nevertheless, in cases where side information is available for users or items, content-based approaches can be included in the ensemble RS. Hypergraph learning has the natural capability of modeling the complex relations between different types of entities in a unified hypergraph and therefore is a deliberate choice to construct an ensemble of RSs with different types of information.

\section{Conclusion}
\label{sec6}
We proposed a new ensemble hypergraph learning-based RS. A unified hypergraph can integrate multiple connections between entities (here users and items) and therefore can combine the predictive power of various individual RSs boosting the precision of final recommendation lists. We empirically tested this method on four datasets from different application domains, such as news, music, and movies. The obtained results showed that the hypergraph-based ensemble RS achieves superior performance compared to all the individual models, as well as compared to a weighted hybrid approach that averages individual scores to produce final rankings, in all datasets.

For future work we outline the following directions:

\begin{itemize}
\item \textbf{Beyond accuracy evaluation}: In this paper we only used user-item interactions. Future approaches could include additional information and relevant stakeholders so that fairness~\cite{fairHyper} and diversity~\cite{diversification_gh} are also taken into account.
\item \textbf{Consumption level}: We only captured the binary feedback between users and items. In real applications usually the user feedback is graded~\cite{gharahighehi2019extended} which shows to what extend the user is interested in the item. This graded feedback could be reflected in the hypergraph to model user preferences more precisely.
\item \textbf{Long-term vs short-term preferences}: In some applications such as news~\cite{gharahighehi2020news} and music~\cite{personalized_diversity_accuracy} recommendation tasks, users' short-term preferences play important roles. Session-based RSs have been used to model such user short-term preferences. An ensemble RS could include models for both long-term and short-term preferences. 
\end{itemize}

\subsubsection*{Acknowledgments}
This work was executed within the imec.icon project NewsButler, a research project bringing together academic researchers (KU Leuven, VUB) and industry partners (Roularta Media Group, Bothrs, ML6). The NewsButler project is co-financed by imec and receives project support from Flanders Innovation \& Entrepreneurship (project nr. HBC.2017.0628). The authors also acknowledge support from the Flemish Government (AI Research Program).

\bibliographystyle{splncs04}
\bibliography{lncs.bbl}
\end{document}